\documentclass[fleqn,usenatbib]{mnras}

\usepackage{newtxtext,newtxmath}

\usepackage[T1]{fontenc}

\DeclareRobustCommand{\VAN}[3]{#2}
\let\VANthebibliography\thebibliography
\def\thebibliography{\DeclareRobustCommand{\VAN}[3]{##3}\VANthebibliography}

\usepackage{graphicx}	
\usepackage{amsmath}	
\usepackage{todonotes}

\newcommand{\msun}{\mbox{M$_\odot$}}
\newcommand{\myr}{\mbox{${\rm Myr}$}}
\newcommand{\msunyr}{\mbox{M$_\odot$~yr$^{-1}$}}

\newcommand*\code[1]{{\fontfamily{lmtt}\selectfont #1}}

\title[Formation of star clusters]{Galactic properties that favour star cluster formation: a statistical view}

\author[Berek, Reina-Campos et al.]{Samantha C. Berek,$^{1,2}$\thanks{E-mail: sam.berek@mail.utoronto.ca} Marta Reina-Campos,$^{1,3,4}$\thanks{E-mail: reinacampos@mcmaster.ca}, Gwendolyn Eadie$^{2,5}$ and Alison Sills$^{3}$\\
$^{1}$Equal contribution\\
$^{2}$David A. Dunlap Department of Astronomy \& Astrophysics, University of Toronto, 50 St. George St., Toronto, ON M5S 3H4, Canada\\
$^{3}$Department of Physics \& Astronomy, McMaster University, 1280 Main Street West, Hamilton, L8S 4M1, Canada\\
$^{4}$Canadian Institute for Theoretical Astrophysics (CITA), University of Toronto, 60 St George St, Toronto, M5S 3H8, Canada\\
$^{5}$Department of Statistical Sciences, University of Toronto, 9th Floor, Ontario Power Building, 700 University Ave, Toronto, ON M5G 1Z5, Canada\\
}

\date{Accepted XXX. Received YYY; in original form ZZZ}

\pubyear{2023}

\begin{document}
\label{firstpage}
\pagerange{\pageref{firstpage}--\pageref{lastpage}}
\maketitle

\begin{abstract}
The presence or absence of star clusters in galaxies, and the properties of star cluster populations compared to their host galaxy properties, are important observables for validating models of cluster formation, galaxy formation, and galaxy assembly. In this work, we apply a Bayesian approach to fit two models to data from surveys of young clusters in star forming galaxies. The first model is a logistic regression, which allows us to include galaxies which do not have any young clusters. The second model is a hurdle model, which includes galaxies with zero clusters and also incorporates information about the total mass in the cluster system. We investigate two predictors (star formation rate and total stellar mass in the galaxy) and look at clusters younger than 10 or 100 Myr. We find that in all cases, star formation rate is the better predictor for both the probability of hosting clusters and the total mass in the cluster system. We compare our results to similar models for old globular clusters, and conclude that star cluster formation was more abundant and more efficient at higher redshifts, likely because of the high gas content of galaxies at that time. 
\end{abstract}



\section{Introduction}

Massive star clusters have been used as tracers of galaxy assembly and formation for many decades. In their seminal work, \citet{SearleZinn1978} used the metallicities of Milky Way globular clusters to postulate that our Galaxy was formed from the hierarchical merging of smaller (dwarf) galaxies and gas clouds. More recently, orbital information for Milky Way globular clusters derived from Gaia proper motions \citep[e.g.][]{Helmi2018,Baumgardt2019} combined with cosmological zoom-in simulations of galaxy formation that include star cluster formation and evolution have confirmed and refined our picture of the assembly of Milky Way-mass galaxies \citep[e.g.][]{Kruijssen19b, Kruijssen2020}.

Despite the apparent stochasticity of hierarchical assembly and cluster formation, observations show an extremely tight, linear relationship between the total mass of a globular cluster system and the host galaxy's dark matter halo mass. \citet{Harris2015} first demonstrated the existence of this relation over five orders of magnitude in stellar mass, and more recent work has confirmed the relation \citep{Burkert2020} and/or extended it in galaxy system mass \citep[e.g.][]{Harris2017,Forbes2018,Dornan23}.

The extension of this relation to low galaxy mass is difficult for two reasons. One, it is very hard to measure the total mass of dwarf galaxies, including their halo, because the tracers that are available at Milky Way mass and above are not as well-behaved for dwarfs. The second reason is more statistical -- at these masses, the predicted number of old massive cluster per galaxy is less than one, whereas star clusters exist in integer quantities. Within a sample of dwarf galaxies, some galaxies will have a cluster (or two or a handful) while others do not have any old clusters. The lowest mass dwarf galaxies are themselves comparable to the mass of a single star cluster in the Milky Way. It has not been clear how to handle this stochasticity in both the observations and models. 

Recent work addresses both of these problems \citep{Eadie2022, Berek2023}. Going beyond the galaxy halo mass and linear regressions, they investigated the connection between globular cluster system mass and the stellar mass of the galaxy, which is easily measurable. These studies also explored models that explicitly allow for zero values, thus including in their models dwarf galaxies that do not host clusters. 

\citet{Eadie2022} tested a variety of models using a sample of Local Group galaxies, nearby isolated dwarfs, and Virgo cluster galaxies, focusing on old globular clusters ($\tau>8~$Gyr). They fit a logistic regression to model the probability of the presence or absence of globular clusters, and find that at a galaxy mass of approximately $\log_{10}(M_{\rm star}/\msun) > 6.8$, galaxies have a $50~$per cent chance of hosting an old stellar cluster. They also tested the use of a hurdle model to incorporate estimates of the population mass for galaxies that do have clusters. 

The hurdle model introduced in \citet{Eadie2022} is expanded by \citet{Berek2023}. The authors develop the HERBAL model, a hierarchical Bayesian hurdle model that uses the luminosity data directly and incorporates measurement uncertainties via an errors-in-variable approach. The HERBAL model treats the true galaxy masses as parameters and also includes a random effect to account for extra scatter in the stellar mass - globular cluster system mass relation. The HERBAL model was fit to a sample of Local Group galaxies and finds that galaxies of approximately $\log_{10}(M_{\rm star}/\msun) = {7.2}$ have at least a $50~$per cent chance of hosting globular clusters. This information provides some important constraints on the galactic conditions under which massive clusters can form.

Models of cluster formation have been successful in reproducing the linear relationship between galaxy mass and globular cluster system mass for both massive galaxies \citep[e.g.][]{BoylanKolchin2017, Choksi2019,El-Badry2019,Bastian2020} and for low-mass dwarf galaxies \citep{Chen2023}. In general, these models either confine themselves to old clusters of 10 Gyr or so (equivalent to Milky Way globular clusters) or simply do not make any age cut. The consensus from these models is that the linear correlation is a hallmark of the combination of many physical processes, including cluster formation and hierarchical galaxy assembly. Cluster formation is inferred to proceed in the same way in both central galaxies and in satellites, with a rate that depends only on local gas or galaxy properties, and galaxy mergers over a Hubble time act mostly to add cluster systems together, with a smaller contribution of merger-driven cluster formation. Cluster disruption can also act to sculpt the present-day distribution of old massive star clusters in galaxies.

Studying old cluster systems, however, means that we are affected by both cluster formation and cluster evolution over most of a Hubble time. Even in dwarf galaxies, clusters are subjected to various processes that will lead to mass loss and possible full cluster disruption, such as stellar evolution, two-body relaxation, and tidal shocks \citep[e.g.][]{GnedinOstriker1997,Lamers05,Gieles08}. However, studying very young cluster populations can provide insight, since they are at most minimally affected by mass loss and evolution \citep{Webb2021}. Comparisons between young and old cluster populations, and how those populations depend on their galactic properties, can help distinguish between the cluster formation process and how clusters evolve in low-mass galaxies.  

In this paper, we make use of the statistical formalism developed in \citet{Eadie2022, Berek2023} for old cluster populations in dwarf galaxies, and apply it to young cluster systems from recent surveys of young massive clusters in actively star-forming galaxies. We investigate the relative importance of galaxy mass and present-day star formation rate on the presence or absence of star clusters, and on the total mass in the cluster system. In Section \ref{sec:datamethods}, we describe the data and our statistical methods; in Section \ref{sec:results} we present the results of our analysis, and we discuss the implications in Section \ref{sec:disc}.

\section{Data and Methods}\label{sec:datamethods}

\subsection{Observational data}\label{sub:obsdata}

We obtain our sample of young star cluster populations in nearby star-forming galaxies from three recently published sources: the ANGST survey\footnote{The ACS Nearby Galaxy Survey Treasury} \citep{Cook2012}, the LEGUS survey\footnote{The Legacy ExtraGalactic UV Survey - \href{https://legus.stsci.edu}{https://legus.stsci.edu}, \citet{Calzetti2015}} \citep{Cook2019,Cook2023} and the PHANGS-HST collaboration\footnote{Physics at High Angular resolution in Nearby GalaxieS - \textit{Hubble Space Telescope} - \href{https://phangs.stsci.edu}{https://phangs.stsci.edu}} \citep{Turner2021, Whitmore2021,Deger2022,Lee2022,Thilker2022}. When a galaxy is present in two or more catalogues, we take the latest measurement as these tend to be more complete and have more accurate total fluxes. We define two cluster samples with different age ranges, one with objects younger than $10~\myr$ and the second with objects younger than $100~\myr$. Although the first sample might be more contaminated by unbound associations \citep[see fig. 15 in][]{Brown2021}, it also represents the latest state of star formation in those galaxies. Overall, our sample contains data on 34 star-forming galaxies from the ANGST survey, 23 from the LEGUS survey and 5 from the PHANGS-HST collaboration.

To characterize the young star cluster populations, we note whether galaxies contain young clusters, the number of young clusters per galaxy and the young clusters' total mass in the two age ranges. The ANGST survey already lists these quantities in \citet{Cook2012}. For the LEGUS dwarf galaxies, we calculate these properties using the latest cluster catalogues \citep{Cook2023}. We use the objects classified as class $1$ or $2$ and with valid masses in the catalogues derived using Padova tracks, Milky Way extinction and the CI-based correction method\footnote{We find no correlation between the classification of the clusters and the properties of their galaxy.}. We also apply a mass cut that selects clusters more massive than $\log_{10}(m/\msun)>3.7$ to have complete samples \citep{Cook2023}. Lastly, for the galaxies from the PHANGS-HST collaboration\footnote{The cluster catalogues are available through the PHANGS homepage at the Mikulski Archive for Space Telescopes with doi:\href{10.17909/t9-r08f-dq31}{10.17909/t9-r08f-dq31}}, we calculate these quantities using class $1$ or $2$ objects from the machine learning catalogues. We use these catalogues because they go deeper in magnitude than the human-classified ones. 

We use the stellar mass and the star formation rate (SFR) to describe the current state of the star-forming galaxies hosting young star cluster populations. For the galaxies present in the ANGST sample, we use the stellar galaxy masses from \citet{Weisz2011}, and the colour-magnitude based SFRs for the last $100~\myr$ from \citet{Cook2012}. The galactic properties of the LEGUS dwarf galaxies are taken from \citet{Cook2019} and \citet{Cook2023}. The stellar masses for the LEGUS dwarf galaxies are computed using $3.6~\mu{\rm m}$ maps from \textit{Spitzer} \citep{Cook2014}, and the SFRs are based on FUV fluxes \citep{Cook2023}. These SFRs roughly correspond to stellar populations of ages of $1$--$100~\myr$ old, and they are within a factor of $2$ of the SFR determined from either $H_{\alpha}$ emission or resolved stellar populations \citep{Cook2023}. Finally, the properties of the PHANGS-HST galaxies are listed in \citet{Leroy2021} and calculated from a variety of sources. The stellar masses are computed from high-quality IRAC $3.6~\mu{\rm m}$ \citep {Sheth2010} and WISE1 $3.4~\mu{\rm m}$ maps \citep{Leroy2019}, and the SFRs are a linear combination of GALEX FUV and WISE4 $22~\mu{\rm m}$ light.

\begin{figure}
\centering
\includegraphics[width=\hsize,keepaspectratio]{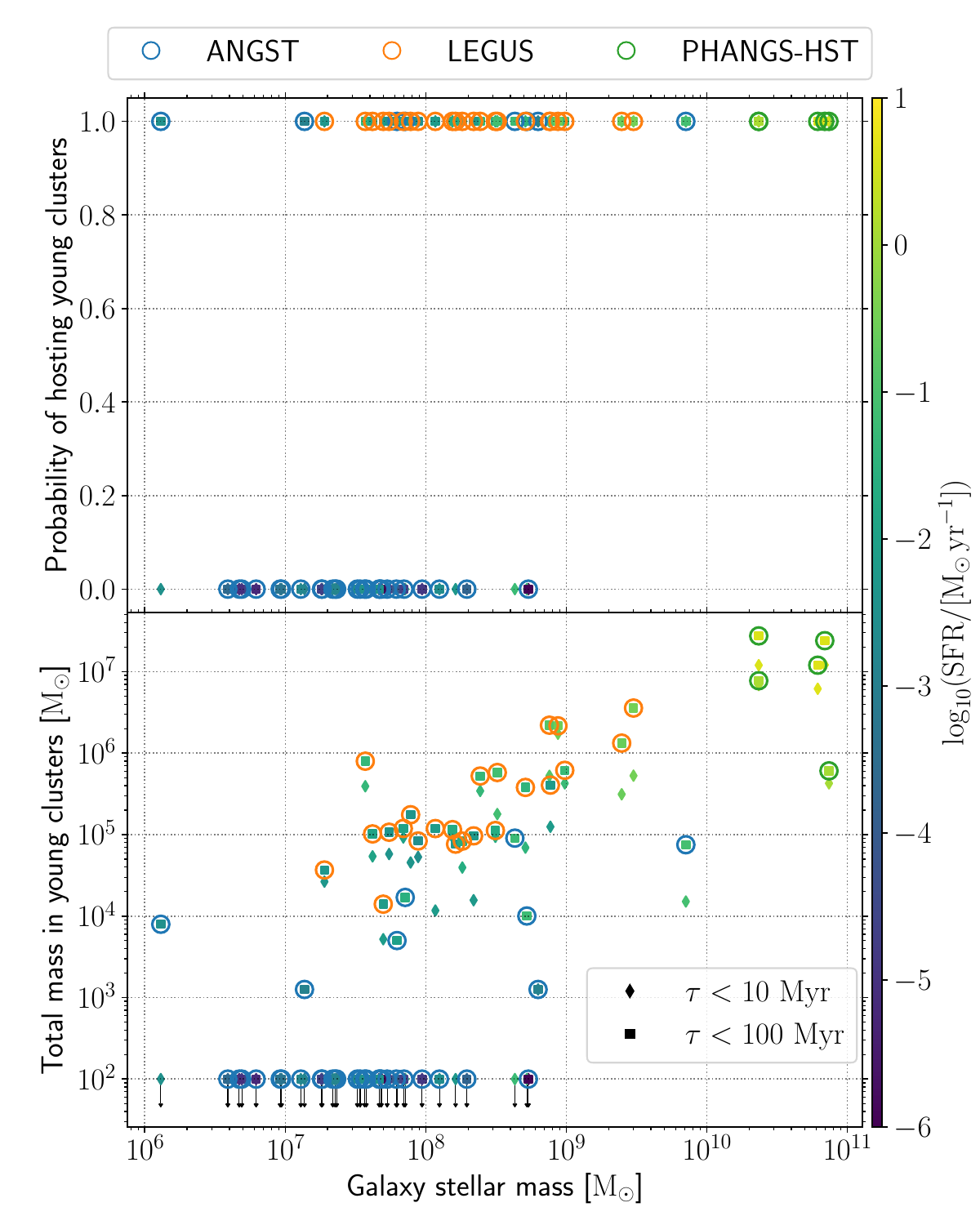}
\caption{\label{fig:clusters-obssample} (\textit{Top}) Flag indicating whether the galaxy hosts young clusters as a function of their galaxy stellar mass.  (\textit{Bottom}) Total mass of young star cluster populations as a function of their host galaxy stellar mass. Galaxies without young clusters are shown as upper limits at $M_{\rm YCs} = 100~\msun$. Data points show cluster populations of two different age ranges as indicated in the legend, and are colour-coded by the SFR of their galaxy. The circles around the points indicate to which catalogue the data belongs.}
\end{figure}

The final sample is shown in Fig.~\ref{fig:clusters-obssample}; the top panel shows the binary response variable (whether or not a galaxy has young clusters) as a function of galaxy stellar mass, while the bottom panel shows the continuous response variable (total mass in young clusters) as a function of galaxy stellar mass. The colours of the points (along with a colour bar on the right-hand side of the figure) indicate the SFR. We see that massive ($M_{\rm star}>10^9~\msun$) galaxies always host young star clusters, whereas low-mass ($M_{\rm star}<10^7~\msun$) systems do not. For galaxies with stellar masses between $M_{\rm star}\sim10^7$--$10^9~\msun$, the existence of clusters seems stochastic: about half of the galaxies contain young star clusters, and the other half do not. When considering the SFR of the galaxy, galaxies with $\rm SFR \sim 10^{-2}~\msunyr$ have a $50~$per cent probability of having young clusters. As long observed for the old globular cluster populations \citep[e.g.][]{Peng08}, the total mass encompassed by the young star cluster population increases towards more massive galaxies. However, the relation shows significant scatter for galaxies with stellar masses between $M_{\rm star}\sim10^7$--$10^9~\msun$, suggesting that cluster formation might be a highly stochastic process in these environments.

\subsection{Statistical Methods}\label{sub:statmethods}
\subsubsection{Logistic Regression}

We begin by looking at the relationship between galaxy mass, SFR, and the presence of young star clusters. For this we use logistic regression, which models the probability of a binary success variable - in this case, the presence of young clusters in a galaxy - based on one or multiple predictor variables. 

The simplest case of the logistic model has a single predictor variable, the value of which determines the probability of the response variable being 1 or 0. For a single predictor model, the probability of success (i.e., containing at least one cluster) is modeled as
\begin{equation}
    \mathbf{p} = \{1+ \exp{\left[-(\beta_0+\beta_1x)\right]}\}^{-1}
\end{equation}
where $x$ is the predictor variable and $(\beta_0,\beta_1)$ are fit coefficients. We use both $\log_{10}{\rm SFR}$ and $\log_{10}{M_{\rm star}}$ separately as predictor variables in this single-predictor case. 

It is straightforward to incorporate multiple predictors into a logistic regression linearly. In a two predictor model, for example, some linear combination of the two predictors $(x_1,x_2)$ (e.g., star formation rate and stellar mass) predicts the presence or absence of young clusters in a galaxy.
In the case where $x_1$ and $x_2$ are assumed not to interact or depend on one another, the second predictor is linearly added to the first as
\begin{equation}
    \mathbf{p} = \{1+ \exp{\left[-(\beta_0+\beta_1x_1+\beta_2x_2)\right]}\}^{-1}
\end{equation} If we add interaction, the probability of success becomes
\begin{equation}
    \mathbf{p} = \{1+ \exp{\left[-(\beta_0+\beta_1x_1+\beta_2x_2+\beta_3x_1x_2)\right]}\}^{-1}
\end{equation}
where we now have three fit coefficients $(\beta_0,\beta_1,\beta_2)$.

\subsubsection{Lognormal Hurdle Model}

The second relationship that we investigate is between our predictor variables $\log_{10}{\rm SFR}$ and $\log_{10}{M_{\rm star}}$ and the mass of young cluster populations. For this we turn to a lognormal hurdle model. Hurdle models have a continuous response variable, in contrast to the binary response from logistic models. Additionally, hurdle models are optimized for data containing a large percentage of zero values, making them well suited for our galaxy sample which contains many galaxies that do not have young clusters. 

A lognormal hurdle model is a combination of a logistic model and a linear model \citep[see][for a more detailed description of the hurdle model applied to astronomical data]{Berek2023}. The logistic model function, as described in the previous section, determines the probability of success (i.e. that a galaxy has young clusters) based on the predictor. Unlike a logistic model, however, the hurdle model describes the non-zero values through a linear regression instead of simply as a binary. The expectation value is a multiplication of these two functions: 
\begin{equation} \label{equ:hurdle}
E[y]=\frac{1}{1+\exp{\left[-(\beta_0+\beta_1 x)\right]}}\left(\gamma_0+\gamma_1 x\right),  
\end{equation}
with the first term being the inverse logit function and the second term being a linear regression. The linear $\beta$ and $\gamma$ terms can be expanded by adding more coefficients to incorporate multiple predictors in the same way as is demonstrated for the logistic function above. 

Uncertainties are incorporated into the hurdle model for both the predictor and response variables via an errors-in-variables model \citep{Berek2023}. For the predictors - $\log_{10}{\rm SFR}$ and $\log_{10}{M_{\rm star}}$ - a normal error distribution is assumed. The measured values of the galaxy properties are considered to be a sample drawn from a normal distribution centered on the true value of the property with a standard deviation equal to the measurement uncertainty: 
\begin{equation}
    x_{\rm meas} \sim \mathcal{N}(x_{\rm true}, \sigma_{\rm meas}).
\end{equation}
In the above, $x_{\rm true}$ is the (unknown) true value of the galaxy property, and $x_{\rm meas}$ and $\sigma_{\rm meas}$ are the measured value and error, respectively.

Uncertainties in the response variable $M_{\rm YCs}$ are built into the linear regression portion of the hurdle model, again assuming a normal distribution of errors. This also assumes that the zero and non-zero populations are entirely separate, as the uncertainties in $y$ only apply to the non-zero portion of the model. 

Along with measurement uncertainty, we follow the conclusions of \citet{Berek2023} and add a parameter for unmeasured scatter $\sigma$ to the overall uncertainty. This scatter term encompasses deviations from linearity and normality, and consists of any intrinsic or otherwise unmeasured sources of uncertainty or scatter in the data. 

\subsubsection{Bayesian methodology}

The logistic and hurdle models were implemented using a Bayesian framework and run using Stan and its RStan interface \citep{stan, rstan}. Stan uses an optimized version of Hamiltonian Monte Carlo (HMC) called a no-U turn sampler (NUTS), which is a gradient-based method of sampling \citep{nuts, hmc}. HMC sampling avoids the random walk behavior of Markov Chain Monte Carlo (MCMC) sampling, which can be prohibitively slow for models with large numbers of parameters. Instead, HMC walks along the peaks of the posterior, exploring only the areas of parameter space that are probable and therefore sampling much more efficiently. 

As in \citet{Berek2023} and \citet{Eadie2022}, we do not have strong prior knowledge of our parameter distributions, and so we leave our priors relatively broad. For the logistic models using SFR as the predictor variable we use priors of: 
\begin{align*}
    \beta_0 &\sim \mathcal{N}(10,5)\\
    \beta_1 &\sim \mathcal{N}(3,5)  
\end{align*}
where $\mathcal{N}(\mu,\sigma)$ are normal distributions of mean $\mu$ and standard deviation $\sigma$. For the logistic models using stellar mass as a predictor, we use priors of: 
\begin{align*}
    \beta_0 &\sim \mathcal{N}(-13,5)\\
    \beta_1 &\sim \mathcal{N}(1,5).  
\end{align*}

\begin{figure*}
\centering
\includegraphics[width=\hsize,keepaspectratio]{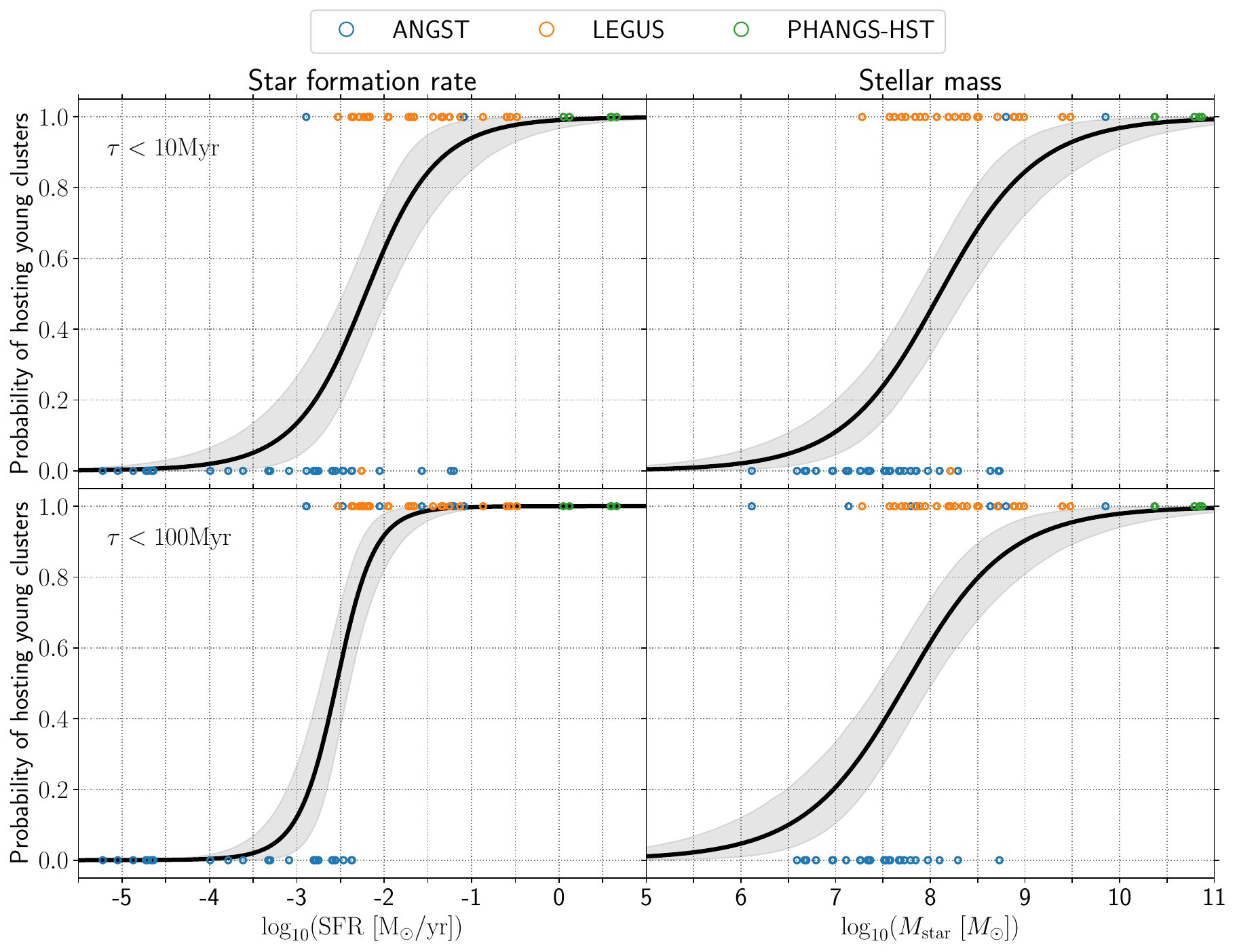}
\caption{\label{fig:logistic-regression-single} Predicted probability of hosting a young star cluster population from the logistic regression models. These models assume a single predictor: the global star formation rate (\textit{left-hand column}) and the galaxy stellar mass (\textit{right-hand column}). Top and bottom panels correspond to cluster populations younger than $10~\myr$ and $100~\myr$, respectively. Different observational samples correspond to the markers as indicated in the legend. The black solid line with the grey shaded region indicate the mean best-fit and the $90~$per cent credible interval, respectively.}
\end{figure*}

The hurdle models require priors for five model parameters: the two logistic $\beta$ parameters, the two linear $\gamma$ parameters, and the unmeasured scatter parameter $\sigma$. For the models using SFR as a predictor, we use priors of:
\begin{align*}
    \beta_0 &\sim \mathcal{N}(10,5)\\
    \beta_1 &\sim \mathcal{N}(3,5)\\
    \gamma_0 &\sim \mathcal{N}(6,2)\\
    \gamma_1 &\sim \mathcal{N}(1,2)\\
    \sigma &\sim \log\mathcal{N}(0,0.5),
\end{align*}
and for the models using stellar mass as a predictor we use priors of:
\begin{align*}
    \beta_0 &\sim \mathcal{N}(-13,5)\\
    \beta_1 &\sim \mathcal{N}(-1,5)\\
    \gamma_0 &\sim \mathcal{N}(0,2)\\
    \gamma_1 &\sim \mathcal{N}(1,2)\\
    \sigma &\sim \log\mathcal{N}(0,0.5).  
\end{align*}

\subsubsection{Model comparison: leave-one-out cross validation}
To compare between versions of our models using different or multiple predictor variables, we use leave-one-out cross validation \citep[LOO-CV;][]{Vehtari2017}. LOO-CV evaluates how good a model is at estimating data not used during the model's training. This allows for numerical comparison of different versions of a model. 

The procedure for LOO-CV consists of removing one point from the dataset and then refitting the model with the remaining data. The expected log predictive density (elpd), which is similar to the log likelihood but based on data not used in training the model, is then calculated for the removed point. This gives a measure of how likely the external point is based on the model. This procedure is repeated for each point in the original dataset, and the elpd values are combined to give an overall measure of how good the model is at predicting data. Higher elpd values indicate a more likely model. However, these values are often multiplied by $-2$ so that they are on the same scale as other commonly used model comparison methods such as the Akaike information criterion (AIC) and Bayesian information criterion (BIC). We follow this convention, meaning that lower values indicate a better model for our data.

\section{Results}\label{sec:results}

We present the best-fit models to describe the probability of a given galaxy to host a young star cluster population, and to predict the total mass in young star clusters. This analysis allows us to investigate which galactic property drives the formation of young star clusters.

\subsection{Logistic Regression Models}

We assume that the probability of hosting young clusters is driven solely by the global SFR or by the galaxy stellar mass. Under this assumption, we fit logistic regression models with a single predictor to the observational data. We provide the best-fit means in Table~\ref{tab:logreg-coeffs} and we show the models in Fig.~\ref{fig:logistic-regression-single}.

Coefficients that are significantly different from zero indicate that the corresponding predictor is providing information about whether or not the galaxy has young clusters\footnote{Caution must be placed when interpreting the constant coefficients, as these do not have an intuitive meaning.}. Similarly, if a coefficient is not significantly different from zero, then its predictor is not providing information. It is thus important to consider the $95~$per cent confidence intervals given on the estimates of each coefficient. 

\begin{table*}
    \caption{\label{tab:logreg-coeffs} Best-fit coefficients of logistic regression models predicting the probability of a given galaxy to host a population of young star clusters. The coefficients are means of the posterior distributions. We apply the models to two samples of cluster populations (younger than $\tau<10$ and $\tau<100~\myr$). The models assume that either the global SFR or the galaxy stellar mass drive the probability of hosting clusters.}
    
\begin{tabular}{lcccccc} 
\hline  
Logistic regression & \multicolumn{3}{c}{$\log_{10}(\rm SFR)$} & \multicolumn{3}{c}{$\log_{10}(M_{\rm star})$} \\ 
& Constant & $\beta_1$ & LOO-CV & Constant & $\beta_1$ & LOO-CV\\ 
\hline 
$\tau < 10~\myr$ & 5.62 & 2.54 & 51.0 & $-$15.97 & 1.97 & 58.7 \\ 
 & (3.07,9.31) & (1.41,4.13) & $\pm$10.1 & ($-$22.68,$-$10.04) & (1.22,2.81) & $\pm$7.2 \\ 
$\tau < 100~\myr$ & 12.19 & 4.79 & 29.7 & $-$14.56 & 1.88 & 63.4\\ 
 & (7.07,18.06) & (2.75,7.13) & $\pm$7.0 & ($-$21.30,$-$8.09) & (1.06,2.75) & $\pm$8.7 \\
\hline 
\textit{Note:}  & \multicolumn{6}{r}{Values in parentheses are $95~$per cent credible intervals} \\ 
  & \multicolumn{6}{r}{$\pm$ values in LOO-CV column are standard errors} \\ 
\end{tabular}

\end{table*}

The probability of hosting young ($\tau<10~\myr$) clusters transitions from 0 to 1 over a wide range in SFRs ($\log_{10}(\rm SFR/\msunyr)={-2.7}$ to $\,-1.6$) and galaxy stellar masses ($\log_{10}(M_{\rm star}/\msun)=7.4$--$8.8$). In contrast, the probability of hosting clusters with ages $\tau<100~\myr$ is tightly constrained by the global SFR of the galaxy. Galaxies with star formation activities above $\log_{10}(\rm SFR/\msunyr)>{-2.5}$ have more than $50~$per cent probability of containing clusters younger than $\tau<100~\myr$. The SFR at which galaxies have a $50~$per cent probability of hosting a young cluster is slightly higher for the young age range. In order to form a young star cluster ($\tau<10~\myr$) after having already formed clusters, galaxies need to sustain a high SFR until the present. However, if their SFR declines after forming a cluster population (e.g.~due to the effects of stellar feedback), then no more clusters form.

A similar picture emerges when examining the best-fit means. These are significantly different from zero for the SFR and the stellar mass, thus indicating that both properties provide predictive information. However, the LOO-CV values suggest that the SFR is a better predictor than the stellar mass for both cluster samples. 

In addition to the models with single predictors, we also consider whether both the SFR and the stellar mass of a galaxy together drive the probability of hosting young clusters. For this, we model the observational data with a logistic regressions model that has multiple predictors, as well as one with an interaction term. Similar to the models with a single predictor, we have no prior knowledge about the priors. We assume that they are well represented by broad Gaussian distributions, but the best-fit coefficients of both models are very sensitive to these assumptions. Regardless of the priors, however, the $\beta_{M_{\rm star}}$ coefficients are not significantly different from zero, and so we conclude that there is no information gain in combining the galaxy properties\footnote{A similar result is obtained for the lognormal hurdle models.}. Thus, we just consider models with single predictors for the rest of our analysis. 

\subsection{Lognormal Hurdle Models}

We run an analogous analysis with the lognormal hurdle models to predict the expected mass in young star clusters. This model has the ability to incorporate uncertainties on both the predictor and the response variables. We do not have systematic uncertainties on the SFR, galaxy stellar mass, or young cluster masses across our observational sample. Therefore, we assume an uncertainty of a factor of two on the young cluster population masses, and a $30~$per cent uncertainty on the galaxy properties. As additional tests, we rerun the analysis for two limiting cases: no measurement uncertainties and inflated uncertainties of a factor of $5$ for the cluster population masses and $50~$per cent on the galaxy properties. We find that within these bracketing limits, the model is robust to changes in the measurement uncertainties.

We investigate the role of the global SFR or the galaxy stellar mass to separately determine the expected mass of the galaxy's young cluster population. The best-fit coefficients are given in Table \ref{tab:hurdle-coeffs}. As in the logistic regression models, the $\beta_1$ coefficients indicate whether the predictor provides information on the probability of galaxies to host young cluster populations. The logistic coefficients of the single predictor hurdle models are similar to those of the logistic regression models, which is to be expected from the fitting algorithm. The $\gamma_1$ coefficients correspond to the slope of the relationship between the predictor and the mass of a galaxy's cluster population for those galaxies that have a non-zero cluster population. Thus, these coefficients can be linked to the rate and efficiency of cluster formation.

\begin{figure*}
\centering
\includegraphics[width=\hsize,keepaspectratio]{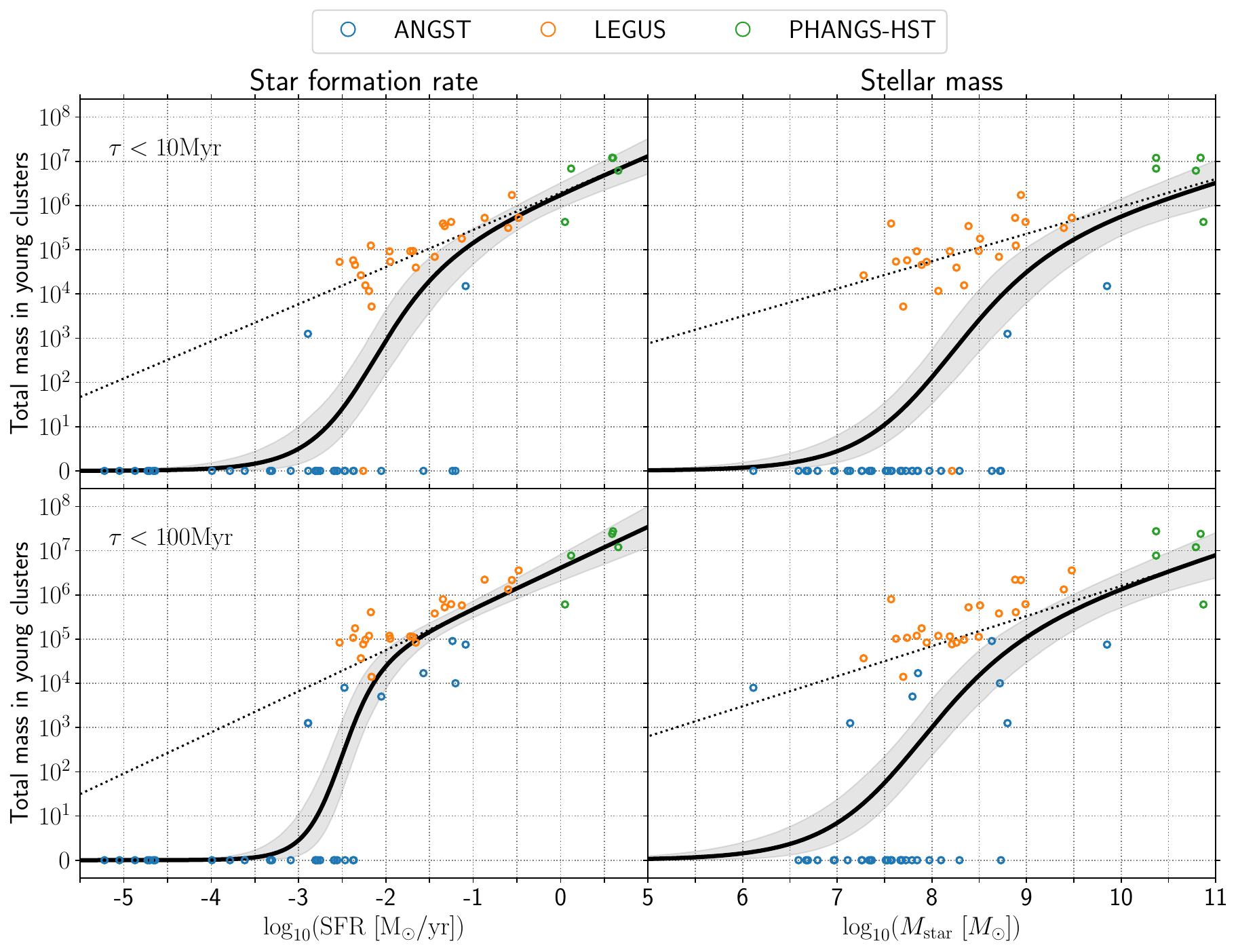}
\caption{\label{fig:hurdle_singlepredictor} Expected masses of young cluster populations from the hurdle model fits. These models assume a single predictor: the global star formation rate (\textit{left-hand column}) and the galaxy stellar mass (\textit{right-hand column}). Top and bottom panels correspond to cluster populations younger than $10~\myr$ and $100~\myr$, respectively. Different observational samples correspond to the markers as indicated in the legend. In each plot, the black line shows the expectation value of the hurdle model. This is a population level value reflecting the average of the zero and non-zero cluster groups at a given value of the predictor, not an expectation value for a single galaxy. The grey shaded areas are the $90~$per cent credible regions, and the dotted lines show just the linear portion of the hurdle model.} 
\end{figure*}

The corresponding best-fit models are shown in Figure \ref{fig:hurdle_singlepredictor}. The solid lines indicate the combined expectation value for galaxies with and without young clusters. Thus, these curves cannot be taken as a prediction for the mass of the young cluster population for any single galaxy. Instead, they show the average size of the cluster population for all galaxies of a given stellar mass or SFR, which includes the galaxies that do not have any clusters at all.

The global SFR of a galaxy is a better predictor of the young cluster population mass than its stellar mass. This result holds for both cluster age ranges, with the youngest range having the lowest LOO-CV value. 

The SFR has a wide transition region (i.e.~the area at which the zero and non-zero populations overlap) for the young age range, $\log_{10}({\rm SFR}/\msunyr)$ between $-3$ to $-1$, but it narrows for the older age range, $\log_{10}({\rm SFR}/\msunyr)$ between $-3$ to $-2$. This result is similar to that found in the logistic regression, and it is a consequence of the similarity between the coefficients. In contrast, the expectation models using the galaxy stellar mass as a single predictor show a wide transition for both age bins. Even in the more massive galaxies, the hurdle model predicts a chance of having no clusters. This is seen by the fact that the hurdle fit (solid black line) does not asymptote to the linear fit (dotted line) until a mass of at least $10^{10}M_\odot$. 

The slopes of the linear portion of the hurdle model contain information on the rate and the efficiency of cluster formation. The slope of the hurdle model using the SFR as a predictor is steeper than that using the galaxy stellar mass. This indicates that cluster population masses increases more rapidly with star formation rate than with galaxy mass. Additionally, the $90~$per cent credible region of the SFR model is more compact than that for the stellar mass model, indicating that there is less model uncertainty about the fit. Overall, we find that the global SFR of a galaxy is a better predictor of the expected total mass in young clusters than its present-day stellar mass.

\begin{table*}
    \caption{\label{tab:hurdle-coeffs} Best-fit coefficients of lognormal hurdle models predicting the probability of a given galaxy to host a population of young star clusters, as well as the expected mass of cluster systems for galaxies that do host clusters. The coefficients are the mean of the posteriors. We apply the models to two samples of cluster populations (younger than $\tau<10$ and $\tau<100~\myr$). The models assume that either the global SFR or the galaxy stellar mass drive the expected mass of cluster systems.}
    


\begin{tabular}{lccccc} 
\hline  
Hurdle model & \multicolumn{5}{c}{$\log_{10}(\rm SFR)$} \\
 & $\beta_0$ & $\beta_1$ & $\gamma_0$ & $\gamma_1$ & LOO-CV \\ 
\hline 
$\tau < 10~\myr$ & 5.84 & 2.62 & 6.29 & 0.84 & 93.0 \\ 
 & (3.09, 9.21) & (1.42, 4.10) & (5.97,6.59) & (0.65,1.02) & $\pm$14.1 \\ 
$\tau < 100~\myr$ & 12.48 & 4.91 & 6.61 & 0.93 & 98.2 \\ 
  & (7.06, 18.96) & (2.75, 7.49) & (6.25,6.98) & (0.72,1.14) & $\pm$13.1 \\ 
\hline  
Hurdle model & \multicolumn{5}{c}{$\log_{10}(M_{\rm star})$} \\
& $\beta_0$ & $\beta_1$ & $\gamma_0$ & $\gamma_1$ & LOO-CV\\ 
\hline 
$\tau < 10~\myr$ & -16.17 & 2.00 & -0.22 & 0.62 & 128.0 \\ 
 & (-23.03, -9.95) & (1.23, 2.86) & (-2.24,1.82) & (0.39,0.84) & $\pm$15.6 \\ 
$\tau < 100~\myr$ & -14.93 & 1.93 & -0.60 & 0.68 & 150.6 \\ 
 & (-21.71, -8.71) & (1.12, 2.80) & (-2.51,1.33) & (0.46,0.91) & $\pm$16.3 \\ 
\hline 
\textit{Note:}  & \multicolumn{5}{r}{Values in brackets are $95~$per cent credible regions} \\ 
  & \multicolumn{5}{r}{$\pm$ values in LOO-CV column are standard errors} \\ 
\end{tabular}


\end{table*}

\section{Summary and Discussion}\label{sec:disc}

In this paper, we used star cluster and galaxy properties from three different surveys of clusters in star-forming galaxies to investigate the relationship between the presence/absence of young clusters and galactic star formation rate and stellar mass. We used logistic regression and lognormal hurdle models in a Bayesian framework, and compared different models using the leave-one-out cross validation method. We find that SFR is a better predictor of both the presence of young star clusters and the total mass in young clusters than the present-day stellar mass of the galaxy. This confirms our expectation that the present-day gas properties of a galaxy will drive the formation of stars and clusters simultaneously. These results also provide some insight into the connection between young clusters and their older, more evolved counterparts. 

\begin{figure}
\centering
\includegraphics[width=\hsize]{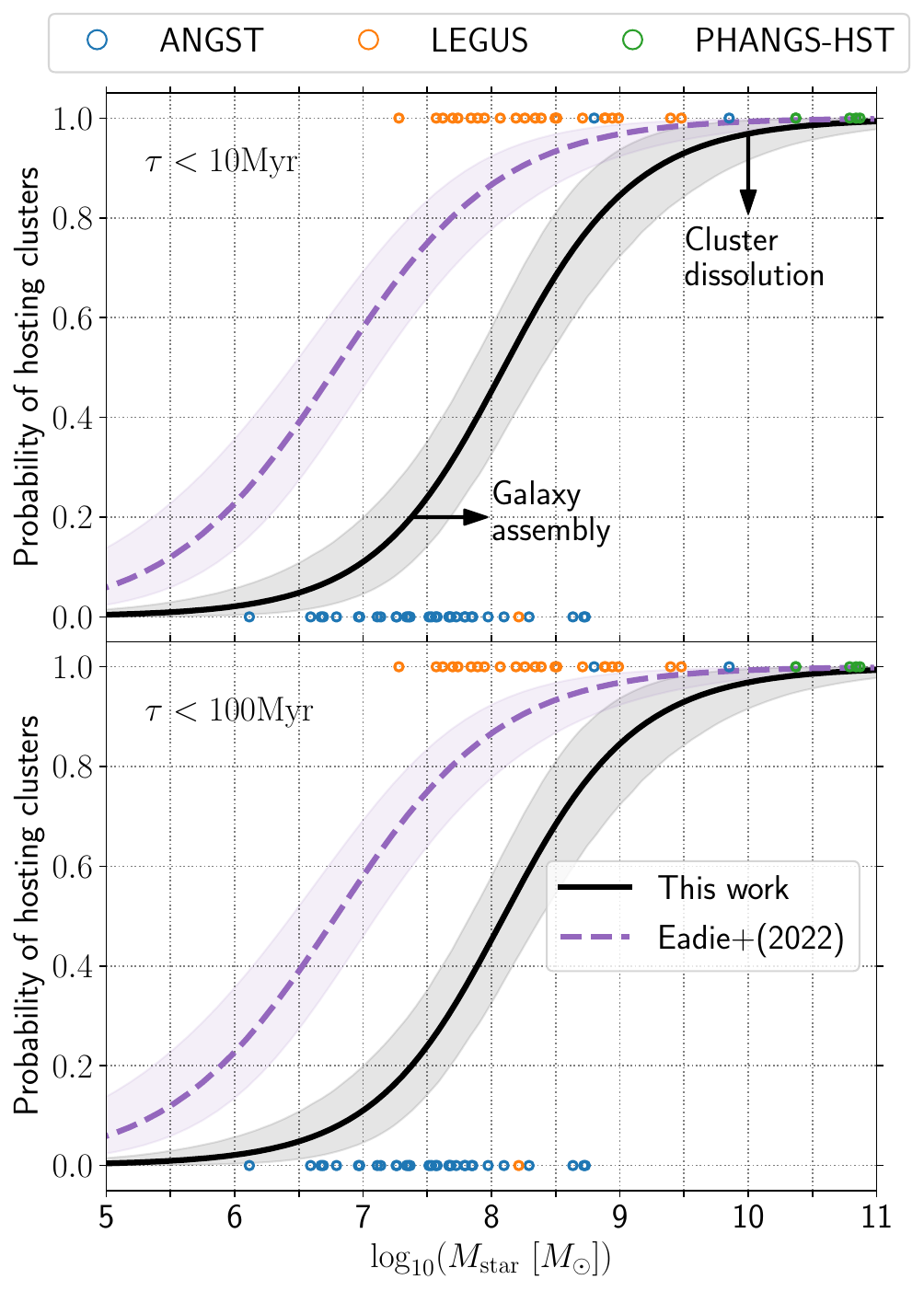}
\caption{\label{fig:logistic_gcmodel} Our logistic regression model for young stellar clusters, using stellar mass as a predictor, is plotted as the black solid line, while the logistic regression for globular clusters from \citet{Eadie2022} is plotted in purple for comparison. The black shaded region corresponds to the $90~$per cent credible interval, and the purple shaded region is the $95~$per cent confidence interval. Markers correspond to the observational data as indicated in the legend. The black arrows show the expected evolution of the young cluster population under the effects of cluster dissolution (where clusters are lost from the galaxy) and galaxy assembly (where the galaxy gains stellar mass over time).
} 
\end{figure}

\begin{figure}
\centering
\includegraphics[width=\hsize]{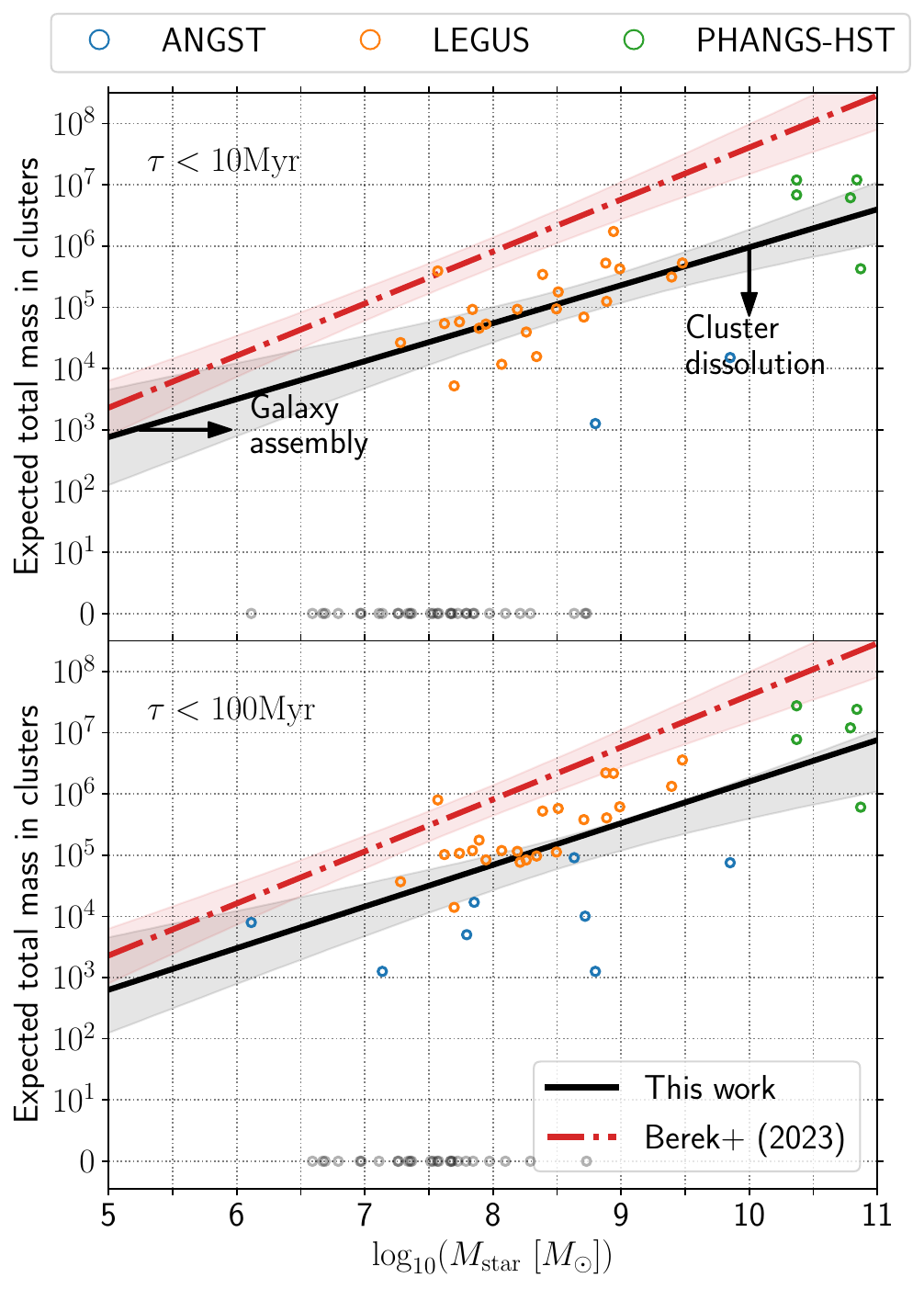}
\caption{\label{fig:hurdle_gcmodel} Hurdle model fits with galaxy stellar mass as a predictor. Our model for young clusters is shown as the sold black line, while the hurdle model fit for globular systems \citep[using a hierarchical mass model;][]{Berek2023} is overplotted as the red dot-dashed line. Shaded regions correspond to the $90~$per cent credible interval. The grey shaded regions shows the 90\% credible regions. Markers correspond to the observational data as indicated in the legend. Grey markers indicate galaxies without a young star cluster population. The black arrows show the expected evolution of the young cluster population under the effects of cluster dissolution (where clusters are lost from the galaxy) and galaxy assembly (where the galaxy gains stellar mass over time).} 
\end{figure}

Similar statistical methods have been used to look at old globular cluster systems in galaxies of a wide range of masses. The probability of hosting clusters is represented in Figure \ref{fig:logistic_gcmodel} for young clusters (black line) from this work, and for globular clusters (purple line) from \citet{Eadie2022}, using the logistic regression models with the galaxy stellar mass as the predictor. We find that galaxies host old clusters at lower stellar masses than young, star-forming galaxies; i.e.~low-mass galaxies in the past could form and retain clusters, while local galaxies of the same mass are not currently forming clusters. 

The linear portion of the expected total cluster population mass from the hurdle model is shown in Figure \ref{fig:hurdle_gcmodel} for young clusters (black line) along with that for old globular clusters (ages greater than 8 Gyr) from \citet{Berek2023}. These lines use the present-day galaxy stellar mass as the predictor for the hurdle model. The data used for the \citet{Berek2023} fit includes cluster masses down to $10^2$ \msun, but the linear portion of that model is not affected by such small systems. Our model for the young clusters has a similar transition region but a shallower slope than the globular cluster model. In other words, older galaxies formed and retained more mass in clusters than galaxies at the present day, and this disparity increases with increasing galaxy mass. 

The black arrows in both figures demonstrate the expected evolution of the young cluster system under two main physical processes. First, clusters do not maintain their birth masses over all time, but lose mass due to stellar evolution, two-body relaxation, and tidal shocks \citep[e.g.][]{GnedinOstriker1997,Lamers05,Gieles08}. In many cases, clusters will be entirely disrupted. Therefore, both the total mass in clusters and the probability of hosting a cluster should decrease with time at constant galaxy mass. Secondly, the stellar mass in galaxies increases with time \citep{Madau2014}. The total stellar mass of a galaxy at the time that it was forming globular clusters (8-10 Gyr ago) should be less than the present day, unless the star formation rate had a sharp and unexpected truncation to zero immediately after the clusters formed. Therefore, the black line should move to the right in these plots. In both ways of modelling the observational data, Figs.~\ref{fig:logistic_gcmodel} and \ref{fig:hurdle_gcmodel}, this expected evolution would move the young population even further from the old globular cluster population, rather than towards it. 
 
These comparisons suggest that it was much easier to form clusters at high redshift. This result is in agreement with many observational and theoretical studies \citep[e.g.][]{PortegiesZwart2010,Forbes2018b,Elmegreen2018, Adamo20}. If the old cluster population was just the evolved version of the current young cluster population, than the globular cluster fits (purple/red lines) would lie below and/or to the right of the young cluster fits (black lines), as described by the arrows. However, this is not the case. Lower-mass galaxies formed clusters 8-10 Gyr ago, and galaxies of all masses have more mass in old clusters than in young ones. At the same time, our analysis of young cluster populations suggests that there is a stronger connection between current star formation rate (i.e. a current gas property) than galaxy stellar mass. Taken together, we conclude that the gas properties of galaxies at early times were more suitable for cluster formation than at the present day. The cluster formation efficiency (defined as the fraction of stars that formed in bound clusters) was higher. This is consistent with our understanding that gas-rich galaxies tend to have higher density gas, resulting in higher star formation rates, higher cluster formation efficiencies, and can form cluster populations that extend to higher initial cluster masses \citep[e.g.][]{Adamo20}.  

The connection between gas properties, young cluster properties, and the subsequent evolution of the cluster system seems fairly straightforward and consistent with our results in this work. However, we are still left with a puzzle -- namely, the extremely tight and extremely linear relationship between the globular cluster system mass and the galaxy dark matter halo mass that is observed over many orders of magnitude in halo mass \citep[e.g.][]{Harris2015, Forbes2018}. This relationship is not obviously a function just of cluster formation physics or even just of cluster evolution. Galaxy assembly must play an important role. One open observational approach that may shed more light on this is the relationship between cluster populations (young and old) and galaxy dark matter mass at the lowest galaxy masses. Of course, determining the total mass of very low-mass galaxies is extremely challenging and has large uncertainties. But if those masses could be determined at high precision for a large number of galaxies \citep[e.g.][]{Oh2015}, then a hurdle analysis using both young and old clusters with halo mass as the predictor would be critical to learn if the clusters are more dependent on the halo (i.e. galaxy)  properties or the star formation rate (i.e. gas properties). 

\section*{Acknowledgements}
The authors would like to thank Joshua Speagle for his insight during many conversations about statistical methods and tests. MRC gratefully acknowledges the Canadian Institute for Theoretical Astrophysics (CITA) National Fellowship for partial support; this work was supported by the Natural Sciences and Engineering Research Council of Canada (NSERC). AIS and GME acknowledges funding from the Natural Science and Engineering Research Council of Canada.

\textit{Software}: This work made use of the following \code{Python} packages: \code{Jupyter Notebooks} \citep{Kluyver16}, \code{Numpy} \citep{numpy-harris20}, \code{Pandas} \citep{pandas_allversions} and \code{Scipy} \citep{Jones01}, and all figures have been produced with the library \code{Matplotlib} \citep{Hunter07}. The comparison to observational data was done more reliably with the help of the WEBPLOT- \code{DIGITIZER}\footnote{\href{https://apps.automeris.io/wpd/}{https://apps.automeris.io/wpd/}} webtool.

Models were run using the Stan Modeling Language \citep{stan} and R Statistical Software Environment \citep{baser}, as well as the following R packages: \texttt{rstan} \citep{rstan}, \texttt{plotrix} \citep{plotrix}, \texttt{ggplot2} \citep{ggplot}, \texttt{latex2exp} \citep{rlatex}, \texttt{bayesplot} \citep{bayesplot}.

\section*{Data Availability}

All data used in this article is available in the original references and we provide a table with the observational data in Appendix~\ref{app:obsdata}.



\bibliographystyle{mnras}
\bibliography{bib_file} 




\appendix

\section{Table of young star clusters and their host galaxies}\label{app:obsdata}

We provide in Table~\ref{tab:obsdata} the data used in this work. The three observational datasets from which we gather the data are the ANGST survey \citep{Cook2012}, the LEGUS survey \citep{Cook2019,Cook2023} and the PHANGS-HST collaboration \citep{Turner2021,Whitmore2021,Deger2022,Lee2022,Thilker2022}. 

\begin{table*}
    \caption{\label{tab:obsdata} Observational data used in this work. Columns indicate the name of the galaxy, its global star formation rate and stellar mass, as well as flags indicating whether the galaxy contain young star clusters and their total mass for two age ranges, and the observational dataset from which the data was taken (see Sect.~\ref{sub:obsdata} for more details). Galaxies are ordered by increasing stellar mass.}
    \begin{tabular}{lccccccc}

   Galaxy & \multicolumn{2}{c}{Galaxy properties} &  
   \multicolumn{2}{c}{Clusters $\tau<10~$Myr} & \multicolumn{2}{c}{Clusters $\tau<100~$Myr} & Sample\\
   
    & SFR [M$_{\odot}$/yr] & M$_{\rm star}$ [M$_\odot$] & Flag & M$_{\rm YSCs}$ & Flag & M$_{\rm YSCs}$ &  \\ \hline
     UGCA292 &      3.35e-03 &      1.30e+06 &                    0 &                 -- &                     1 &            7.94e+03 &       ANGST \\
       KKR03 &      2.42e-04 &      3.90e+06 &                    0 &                 -- &                     0 &                  -- &       ANGST \\
    HS98-117 &      8.91e-06 &      4.70e+06 &                    0 &                 -- &                     0 &                  -- &       ANGST \\
 ESO540-G030 &      2.28e-05 &      4.90e+06 &                    0 &                 -- &                     0 &                  -- &       ANGST \\
    FM2000-1 &      6.00e-06 &      6.20e+06 &                    0 &                 -- &                     0 &                  -- &       ANGST \\
       KKH98 &      4.97e-04 &      9.20e+06 &                    0 &                 -- &                     0 &                  -- &       ANGST \\
       KDG73 &      8.14e-04 &      9.40e+06 &                    0 &                 -- &                     0 &                  -- &       ANGST \\
     UGC8091 &      1.77e-03 &      1.29e+07 &                    0 &                 -- &                     0 &                  -- &       ANGST \\
     UGC9128 &      1.29e-03 &      1.37e+07 &                    0 &                 -- &                     1 &            1.26e+03 &       ANGST \\
     UGCA276 &      1.98e-05 &      1.81e+07 &                    0 &                 -- &                     0 &                  -- &       ANGST \\
        BK3N &      1.57e-03 &      1.83e+07 &                    0 &                 -- &                     0 &                  -- &       ANGST \\
     UGC4459 &      5.17e-03 &      1.90e+07 &                    1 &           2.65e+04 &                     1 &            3.68e+04 &       LEGUS \\
     UGC8833 &      1.56e-03 &      2.18e+07 &                    0 &                 -- &                     0 &                  -- &       ANGST \\
      M81dwA &      1.66e-03 &      2.26e+07 &                    0 &                 -- &                     0 &                  -- &       ANGST \\
       KKH37 &      4.71e-04 &      2.32e+07 &                    0 &                 -- &                     0 &                  -- &       ANGST \\
     UGC8508 &      2.77e-03 &      3.26e+07 &                    0 &                 -- &                     0 &                  -- &       ANGST \\
    ArpsLoop &      2.54e-03 &      3.41e+07 &                    0 &                 -- &                     0 &                  -- &       ANGST \\
     UGC8651 &      3.41e-03 &      3.67e+07 &                    0 &                 -- &                     0 &                  -- &       ANGST \\
     UGC5340 &      4.50e-02 &      3.71e+07 &                    1 &           3.92e+05 &                     1 &            7.97e+05 &       LEGUS \\
     NGC3741 &      4.22e-03 &      3.77e+07 &                    0 &                 -- &                     0 &                  -- &       ANGST \\
     UGCA281 &      1.12e-02 &      4.18e+07 &                    1 &           5.41e+04 &                     1 &            1.02e+05 &       LEGUS \\
     UGC5442 &      1.00e-32 &      4.66e+07 &                    0 &                 -- &                     0 &                  -- &       ANGST \\
     UGC8760 &      4.29e-03 &      4.69e+07 &                    0 &                 -- &                     0 &                  -- &       ANGST \\
     UGC5428 &      1.00e-32 &      4.83e+07 &                    0 &                 -- &                     0 &                  -- &       ANGST \\
     UGC5139 &      6.85e-03 &      4.98e+07 &                    1 &           5.20e+03 &                     1 &            1.40e+04 &       LEGUS \\
     UGCA133 &      1.87e-05 &      5.31e+07 &                    0 &                 -- &                     0 &                  -- &       ANGST \\
      IC4247 &      4.22e-03 &      5.48e+07 &                    1 &           5.78e+04 &                     1 &            1.07e+05 &       LEGUS \\
       KDG61 &      1.02e-04 &      6.17e+07 &                    0 &                 -- &                     0 &                  -- &       ANGST \\
     UGC9240 &      8.86e-03 &      6.24e+07 &                    0 &                 -- &                     1 &            5.01e+03 &       ANGST \\
       IC559 &      1.10e-02 &      6.92e+07 &                    1 &           9.23e+04 &                     1 &            1.19e+05 &       LEGUS \\
       DDO78 &      2.18e-05 &      7.00e+07 &                    0 &                 -- &                     0 &                  -- &       ANGST \\
     UGC5336 &      2.69e-02 &      7.14e+07 &                    0 &                 -- &                     1 &            1.70e+04 &       ANGST \\
     UGC7242 &      4.43e-03 &      7.80e+07 &                    1 &           4.55e+04 &                     1 &            1.75e+05 &       LEGUS \\
      UGC685 &      2.95e-03 &      8.82e+07 &                    1 &           5.32e+04 &                     1 &            8.36e+04 &       LEGUS \\
  LEDA166101 &      1.34e-05 &      9.41e+07 &                    0 &                 -- &                     0 &                  -- &       ANGST \\
     NGC5238 &      6.41e-03 &      1.17e+08 &                    1 &           1.17e+04 &                     1 &            1.19e+05 &       LEGUS \\
     NGC4163 &      2.58e-03 &      1.25e+08 &                    0 &                 -- &                     0 &                  -- &       ANGST \\
     NGC5477 &      1.90e-02 &      1.55e+08 &                    1 &           9.27e+04 &                     1 &            1.15e+05 &       LEGUS \\
      UGC695 &      5.50e-03 &      1.63e+08 &                    0 &                 -- &                     1 &            7.65e+04 &       LEGUS \\
 ESO486-G021 &      2.20e-02 &      1.82e+08 &                    1 &           3.95e+04 &                     1 &            8.32e+04 &       LEGUS \\
         IKN &      1.64e-04 &      1.96e+08 &                    0 &                 -- &                     0 &                  -- &       ANGST \\
     UGC7408 &      5.82e-03 &      2.19e+08 &                    1 &           1.57e+04 &                     1 &            9.66e+04 &       LEGUS \\
     NGC1705 &      4.71e-02 &      2.44e+08 &                    1 &           3.44e+05 &                     1 &            5.24e+05 &       LEGUS \\
     UGC4305 &      2.04e-02 &      3.13e+08 &                    1 &           9.40e+04 &                     1 &            1.12e+05 &       LEGUS \\
     NGC3274 &      7.40e-02 &      3.23e+08 &                    1 &           1.79e+05 &                     1 &            5.78e+05 &       LEGUS \\
     UGC8201 &      5.79e-02 &      4.31e+08 &                    0 &                 -- &                     1 &            9.06e+04 &       ANGST \\
     NGC3738 &      3.62e-02 &      5.12e+08 &                    1 &           6.92e+04 &                     1 &            3.80e+05 &       LEGUS \\
     NGC2366 &      6.26e-02 &      5.24e+08 &                    0 &                 -- &                     1 &            1.00e+04 &       ANGST \\
      NGC404 &      1.00e-32 &      5.37e+08 &                    0 &                 -- &                     0 &                  -- &       ANGST \\
     UGC5692 &      1.28e-03 &      6.30e+08 &                    1 &           1.26e+03 &                     1 &            1.26e+03 &       ANGST \\
     NGC4485 &      1.35e-01 &      7.59e+08 &                    1 &           5.27e+05 &                     1 &            2.20e+06 &       LEGUS \\
     NGC4248 &      6.72e-03 &      7.71e+08 &                    1 &           1.25e+05 &                     1 &            4.05e+05 &       LEGUS \\
     NGC5253 &      2.77e-01 &      8.73e+08 &                    1 &           1.73e+06 &                     1 &            2.16e+06 &       LEGUS \\
     UGC1249 &      5.58e-02 &      9.76e+08 &                    1 &           4.26e+05 &                     1 &            6.15e+05 &       LEGUS \\
     NGC4656 &      2.51e-01 &      2.48e+09 &                    1 &           3.13e+05 &                     1 &            1.33e+06 &       LEGUS \\
     NGC4449 &      3.31e-01 &      3.01e+09 &                    1 &           5.29e+05 &                     1 &            3.58e+06 &       LEGUS \\
      IC2574 &      8.18e-02 &      7.10e+09 &                    1 &           1.51e+04 &                     1 &            7.51e+04 &       ANGST \\
     NGC1559 &      3.98e+00 &      2.34e+10 &                    1 &           1.20e+07 &                     1 &            2.75e+07 &  PHANGS-HST \\
     NGC3351 &      1.32e+00 &      2.34e+10 &                    1 &           6.85e+06 &                     1 &            7.74e+06 &  PHANGS-HST \\
     NGC1566 &      4.57e+00 &      6.17e+10 &                    1 &           6.16e+06 &                     1 &            1.20e+07 &  PHANGS-HST \\
     NGC3627 &      3.89e+00 &      6.92e+10 &                    1 &           1.21e+07 &                     1 &            2.40e+07 &  PHANGS-HST \\
     NGC1433 &      1.12e+00 &      7.41e+10 &                    1 &           4.26e+05 &                     1 &            6.06e+05 &  PHANGS-HST \\
\end{tabular}
\end{table*}


\bsp	
\label{lastpage}
\end{document}